\documentclass[12pt]{article}

\usepackage{amsmath,amsthm,amsfonts,amssymb,amscd}
\def\qed{{\unskip\nobreak\hfil\penalty50
\hskip2em\hbox{}\nobreak\hfil$\square$
\parfillskip=0pt \finalhyphendemerits=0\par}\medskip}

\def\Tr{{\mathrm {Tr}}}

\def\Tr{{\mathrm {Tr}}}

\newtheorem{theorem}{Theorem}[section]
\newtheorem{lemma}[theorem]{Lemma}
\newtheorem{conjecture}[theorem]{Conjecture}
\newtheorem{corollary}[theorem]{Corollary}
\newtheorem{definition}[theorem]{Definition}

\newtheorem{proposition}[theorem]{Proposition}
\newtheorem{remark}[theorem]{Remark}

\def\res{\!\restriction\!}
\def\PO{{\cal PO}}
\def\A{{\cal A}}

\def\B{{\cal B}}

\def\PI{{\cal PI}}

\def\H{{\cal H}}
\renewcommand{\qed}{\ \hfill $\blacksquare$}

\newcommand{\bdef}{\begin{definition}}
\newcommand{\blem}{\begin{lemma}}
\newcommand{\bprop}{\begin{proposition}}
\newcommand{\bthm}{\begin{theorem}}
\newcommand{\bcor}{\begin{corollary}}
\newcommand{\bconj}{\begin{conjecture}}
\newcommand{\ben}{\begin{equation}}
\newcommand{\een}{\end{equation}}

\newcommand{\ede}{\end{definition}}
\newcommand{\elem}{\end{lemma}}
\newcommand{\eprop}{\end{proposition}}
\newcommand{\ethm}{\end{theorem}}
\newcommand{\ecor}{\end{corollary}}
\newcommand{\econj}{\end{conjecture}}
\newcommand{\brem}{\begin{remark}}
\newcommand{\erem}{\end{remark}}

\newcommand{\ba}{\begin{array}}
\newcommand{\ea}{\end{array}}
\newcommand{\bea}{\begin{eqnarray}}

\title{\huge Some Results On Relative Entropy in Quantum Field Theory}
\author{
{\sc Feng Xu}\footnote{Supported in part by NSF grant DMS-1764157.}\\
Department of Mathematics\\
University of California at Riverside\\
Riverside, CA 92521\\
E-mail: {\tt xufeng@math.ucr.edu}}

\begin{document}
\date{}
\maketitle

\begin{abstract}
We prove that the mutual information for vacuum state as defined by
Araki is finite for  quantum field theory of free fermions on a
Minkowski spacetime of any dimension. In the case of  two
dimensional chiral conformal field theory (CFT) we use our previous
results for the free fermions to show that for a large class of
chiral CFT the mutual information is finite. We also provide a
direct relation between relative entropy and  the index of a
representation of conformal net.

\end{abstract}

\newpage

\section{Introduction}
In the last few years there has been an enormous amount of work by
physicists concerning entanglement entropies in QFT, motivated by
the connections with condensed matter physics, black holes, etc.;
see the references in \cite{Hollands} for a partial list of
references.   However, some very basic mathematical questions remain
open. Often, the mutual information is argued to be finite based on
heuristic physical arguments, and one can derive the singularities
of the entropies from the mutual information by taking singular
limits. But it is not even clear that such mutual information, which
is well defined as a special case of Araki's relative entropy, is
indeed finite. All the heuristic computations such those done in
\cite{Casf} and \cite{Car} take this for granted and these papers
contain  a number of amazing results about the nature of such mutual
information. It is clear that there should be a rich mathematical
theory behind these physical considerations. See \cite{Hol},
\cite{Lon}, \cite{L2}, \cite{LXc}, \cite{LXr} , \cite{OT} and
\cite{Wit} for a partial list of recent mathematical work.

In \cite{LXr} we showed that mutual information for massless free
fermions is finite, and in \cite{LXr} we calculate its value for all
cases. In fact this is the only example where all mutual information
is known (see \cite{Casb} for recent computations in the case of
massless bosons with two intervals). The method in \S3 of \cite{LXr}
uses the explicit resolvent formula for free fermions which
unfortunately is not known in other cases such as free massive
fermions. One of the goals on this paper is to improve on some of
the estimates in \S3 of \cite{LXr} so that we can obtain finiteness
of mutual information for all free fermion theories (cf. Corollary
\ref{free}) without having  the explicit resolvent formula
available. The main results which lead to Corollary \ref{free} are
Theorem \ref{key} and Proposition \ref{smooth}.

We then consider more such finiteness results for chiral CFT in two
dimensions by embedding into free fermions and using monotonicity of
relative entropy. First we show that every lattice contains a finite
index sublattice which can be embedded into free fermions in
Corollary \ref{emb}. As a consequence we show that mutual
information is finite for all conformal net coming from lattices in
Corollary \ref{lat2}. These immediately show that all conformal nets
which can be embedded as a subnet of conformal nets associated with
a lattice, and   with a simply connected Lie group $G$ at level $k$
or so called Wess-Zumino-Witten models, their cosets, orbifolds,
 simple current extensions and combinations of such constructions, the mutual information is always
finite. Our last result Theorem \ref{index} gives a direct relation
between relative entropy and the index of a representation of
conformal net, in a similar spirit to a result in \S4 of \cite{LXr}.

The rest of this paper is structured as follows. After a preliminary
section on von Neumann entropy, Araki's relative entropy,
 we consider the  mutual information in an algebraic quantum field
 theory with split property, and use free fermion theory as an
 example. Then we consider a general problem motivated by
 computations of mutual information in \S\ref{general}, where we
 prove a few keys results such as Thereom \ref{key}, Proposition
 \ref{smooth}. The finiteness of mutual information in free QFT is
 obtained as a consequence in Corollary \ref{free}. In \S\ref{cft}
 we first show that a conformal net $\A_L$ associated with a lattice
 has a finite index subnet which can be embedded into free fermions.
 It follows by monotonicity of relative entropy that mutual
 information for $\A_L$ is finite. From this we derive the
 finiteness of mutual information for a large class of chiral CFTs.
  In the last
subsection we prove Theorem \ref{index}.

 \section{Preliminaries}\label{prelim}

\subsection{Entropy and relative entropy}

von Neumann entropy is the quantity associated with a density matrix $\rho$ on a
Hilbert space $\H$ by
\[
S(\rho) = -\Tr(\rho \log \rho)\ .
\]
von Neumann entropy can be viewed as a measure of the lack of information about a
system to which one has ascribed the state. This interpretation is
in accord for instance with the facts that $S(\rho)\ge0$ and that a
pure state $\rho = |\Psi \rangle \langle \Psi|$ has vanishing von
Neumann entropy.

A related notion is that of the relative entropy. It is defined for
two density matrices $\rho, \rho'$ by \ben  S(\rho,\rho') = \Tr(\rho
\log \rho - \rho \log \rho') \ . \een  Like $S(\rho)$,
$S(\rho,\rho')$ is non-negative, and can be infinite.

A generalization of the relative entropy in the context of von
Neumann algebras of arbitrary type was found by Araki~\cite{arakir}
and is formulated using modular theory. Given two faithful, normal
states $\omega, \omega'$ on a von Neumann algebra $\A$ in standard
form, we choose the vector representatives in the natural cone
$\mathcal{P}^\sharp$, called $|\Omega \rangle,|\Omega' \rangle$ .
The anti-linear operator
 $S_{\omega,\omega'} a|\Omega' \rangle = a^*
|\Omega \rangle$, $a\in \A$, is closable
and one considers again the polar decomposition of its closure
$\bar S_{\omega, \omega'} = J \Delta_{\omega,\omega'}^{1/2}$ . Here $J$ is the
modular conjugation of $\A$ associated with $\mathcal{P}^\sharp$ and $\Delta_{\omega,\omega'} =
 S^*_{\omega, \omega'}\bar S_{\omega, \omega'}$ is the relative modular operator w.r.t. $|\Omega \rangle,|\Omega' \rangle$.
Of course, if $\omega = \omega'$ then $\Delta_{\omega} = \Delta_{\omega,\omega'}$ is the usual modular operator.

A related
object is the Connes cocycle (Radon-Nikodym derivative) defined as
$[D\omega : D\omega']_t = \Delta_{\omega, \psi}^{it} \Delta_{\psi,
\omega'}^{it} \in \A$, where $\psi$ is an arbitrary auxiliary
faithful normal state on $\A'$ . \bdef
 The relative entropy w.r.t. $\omega$ and $\omega'$ is defined by
  \ben\label{drel1}
S(\omega, \omega') = \langle \Omega | \log \Delta_{\omega, \omega'}
\ \Omega\rangle  = \lim_{t \to 0} \frac{\omega([D\omega:D\omega']_t
- 1)}{it} \ , \een $S$ is extended to positive linear functionals that are
not necessarily normalized by the formula
$S(\lambda\omega,\lambda'\omega')=\lambda S(\omega,\omega') +
\lambda \log (\lambda/\lambda')$, where $\lambda, \lambda'>0$ and
$\omega,\omega'$ are normalized. If $\omega'$ is not normal, then
one sets $S(\omega, \omega') = \infty$. \ede

 For a type I algebra $\A = \B(\H)$, states $\omega, \omega'$
correspond to density matrices $\rho, \rho'$. The square root of the relative modular
operator $\Delta_{\omega,\omega'}^{1/2}$ corresponds to $\rho^{1/2}
\otimes \rho^{\prime -1/2}$ in the standard representation of $\A$ on $\H
\otimes \bar \H$; namely $\H
\otimes \bar \H$ is identified with the Hiilbert-Schmidt operators $HS(\H)$ with the left/right multiplication
of $\A$/$\A'$.
 In this representation, $\omega$ corresponds to
the vector state $|\Omega \rangle =\rho^{1/2} \in \H \otimes
\bar{\H}$, and the abstract definition of the relative entropy
in~\eqref{drel1} becomes \ben \langle \Omega | \log \Delta_{\omega,
\omega'} \, \Omega \rangle = \Tr_{\H} \rho^{\frac12}\left(\log\rho
\otimes 1 - 1 \otimes \log \rho'\right)\rho^{\frac12} = \Tr_\H(\rho
\log \rho - \rho \log \rho') \ . \een

\medskip

As another example, let us consider a bi-partite system with Hilbert
space $\H_A \otimes \H_B$ and observable algebra $\A = \B(\H_A)
\otimes \B(\H_B)$. A  normal state $\omega_{AB}$ on $\A$ corresponds to a
density matrix $\rho_{A B}$.  One calls $\rho_A = \Tr_{\H_B}
\rho_{AB}$ the ``reduced density matrix'', which defines a state
$\omega_A$ on $\B(\H_A)$ (and similarly for system $B$).
 The mutual
information is given in our example system by
\ben\label{remu}
S(\rho_{AB},\rho_A\otimes \rho_B ) = S(\rho_A) +S(\rho_B) -
S(\rho_{A B}) \ . \een

A list of properties of relative entropies that will be used later
can be found in \cite{OP} (cf. Th. 5.3, Th. 5.15 and Cor. 5.12
\cite{OP}):

\bthm\label{515} (1) Let $M$ be a von Neumann algebra and $M_1$ a
von Neumann subalgebra of M. Assume that there exists a faithful
normal conditional expectation $E$ of $M $onto $M_1$. If $\psi$ and
$\omega$  are states of $M_1$ and $M$, respectively, then $S(\omega,
\psi\cdot E) = S(\omega\res M_1, \psi) + S(\omega, \omega\cdot E);$
\par

(2) Let be $M_i$ an increasing net of von Neumann subalgebras of $
M$ with the property $ (\bigcup_i M_i)''=M$. Then $S(\omega_1\res
M_i, \omega_2\res M_i)$ converges to $ S(\omega_1,\omega_2)$ where
$\omega_1, \omega_2$ are two normal states on $M$; \par

(3) Let $\omega$ and  $\omega_1$ be two normal states on a von
Neumann algebra $M$. If $\omega_1\geq \mu\omega,  $ then $S(\omega,
\omega_1) \leq \ln \mu^{-1}$;

(4) Let $\omega$ and  $\phi$ be two normal  states on a von Neumann
algebra $M$, and denote by  $\omega_1$ and  $\phi_1$ the
restrictions of   $\omega$ and  $\phi$ to  a von Neumann subalgebra
$M_1\subset M$ respectively. Then $S(\omega_1, \phi_1)\leq S(\omega,
\phi)$.
\ethm
For type  $\mathrm{III}$  factors, the von Neumann entropy is always
infinite, but we shall see that in many cases mutual information is
finite.

Let us describe the setting where the relative entropy we are interested in computing. We consider the formulation of algebraic quantum field theory on a $D=d+1$
 dimensional Minkowski spacetime (cf. \cite{Haa}). Let $DO$ be an open subset of space time such that the closure of $DO$ is compact.
Let $\A(DO)$ be the algebra of observable associated with $DO$, and
$\omega$ the vector state given by the vacuum vector. For simplicity
we will assume that $DO$ is the double cone generated by an open set
$O$ on the time zero slice  ${\Bbb R}^d$. We shall assume that $O$
has smooth boundary and the closure of $O$ in ${\Bbb R}^d$ is
compact. Slightly abusing our notation we denote $\A(DO)$ simply by
$\A(O)$ . $O_1$ and $O_2$ are disjoint if $\bar{O_1}\cap \bar{O_2}
=\emptyset.$ Denote by $\PO$ the set which consists of finite union
of disjoint $O$s. Let $O_1, O_2\subset \PO$, with $\bar{O_1}\cap
\bar{O_2} =\emptyset.$  Let $\omega_1, \omega_2$ be the restriction
of $\omega$ to $\A(O_1), \A(O_2)$ respectively.

We shall assume that our theory is split (cf. \cite{Buc} for Bosonic
case and \cite{Sum} for fermionic case), which means that
$\omega_1\otimes \omega_2$, which is defined on elements of the form
$xy, x\in \A(O_1), y\in\A(O_2)$ by $\omega_1\otimes \omega_2(xy)=
\omega_1(x)  \omega_2(y),$ extends to a normal faithful state of the
von Neumann algebra generated by $\A(O_1)$ and $\A(O_2).$ The basic
quantity we are interested in is the relative entropy (also called
mutual information) $S(\omega, \omega_1\otimes \omega_2).$

As an example let us  consider  chiral free fermion CFT as discussed
in details in \S3 of \cite{LXr}. We will describe the formula for
mutual information which is proved in Th. 3.18 of  \cite{LXr}, and
refer reader to  \S3 of \cite{LXr} for more details.

Let $H$ denote the Hilbert space $L^2(S^1; \mathbb C^r)$ of
square-summable $\mathbb{C}^r$-valued functions on the circle.

Fix $I_i\in \PI, i=1,2$, and $I_1, I_2$ disjoint, that is $\bar
{I_1} \cap \bar {I_2} = \emptyset$, and $I=I_1\cup I_2$. Denote by
$\A_r$ the graded conformal net associated with $r$ chiral free
fermions. We will write the normal faithful state $\omega_1\otimes_2
\omega_2$ with graded tensor product in \S3 of \cite{LXr} simply as
$\omega_1\otimes\omega_2,$ and the mutual information we are
interested is now $S_{\A_r}(\omega,\omega_1\otimes \omega_2).$

The vacuum state $\omega$ on $\A_r(I)$ is  a quasi-free state as
studied by Araki in \cite{arakif}. To describe this state, it is
convenient to use Cayley transform $V(x) = (x-i) / (x + i)$, which
carries the (one point compactification of the) real line onto the
circle and the upper half plane onto the unit disk. It induces a
unitary map $$ {\displaystyle Uf(x)=\pi ^{-{\frac
{1}{2}}}(x+i)^{-1}f(V(x))}  $$ of $L^2(S^1, \mathbb{C}^r)$ onto
$L^2(\mathbb{R},\mathbb{C}^r )$. The operator $U$ carries the Hardy
space on the circle onto the Hardy space on the real line . We will
use the Cayley transform to identify intervals on the circle with
one point removed to intervals on the real line. Under the unitary
transformation above, the Hardy projection on $L^2(S^1,
\mathbb{C}^r)$ is transformed to the Hardy projection on
$L^2(\mathbb{R},\mathbb{C}^r )$ given by
$$
Pf(x) = \frac{1}{2}f(x) + \int \frac{i}{2 \pi} \, \frac{1}{(x-y)}
f(y)dy \ ,
$$
where the singular integral is (proportional to) the Hilbert transform.

We write the kernel of the above integral transformation as $C$:
\begin{equation}
C(x,y)=\frac{1}{2} \delta(x-y) - \frac{i}{2 \pi} \,
\frac{1}{(x-y)}\, \ .
\end{equation}
The quasi free  state $\omega$ is determined by $$\omega
\big(a(f)^*a(g)\big)= \langle g, P f\rangle.$$ Slightly abusing our
notations, we will identify  $P$ with its kernel $C$ and simply
write  $$\omega \big(a(f)^*a(g) \big)= \langle g, C f\rangle.$$  $C$ will be
called   {\it covariance operator}.

\bdef\label{sigma}

Let $\mathbf{P}_i$ be projections from $L^2(I,\mathbb{C}^r)$ onto
$L^2({I_i},\mathbb{C}^r),$ and $C_i= \mathbf{P}_iC \mathbf{P}_i,
i=1,2$.

Let
\begin{multline*}
\sigma_C = \mathbf{P_1}\big (C\ln C +(1-C)\ln
(1-C)\big)\mathbf{P_1}- \big(C_1
\ln C_1 +  (\mathbf{P_1}-C_1)\ln( \mathbf{P_1}-C_1)\big)+ \\
\mathbf{P_2}\big(C\ln C +(1-C)\ln (1-C)\big)\mathbf{P_2}- \big(C_2
\ln C_2+(\mathbf{P_2}-C_2)\ln( \mathbf{P_2}-C_2)\big)
\end{multline*}
 and $\sigma_{C_p}$ be
the same as in the definition of $\sigma_C $ with $C$ replaced by
$C_p= pCp,$ if $p$ is a projection commuting with $\mathbf{P}_1.$

\ede

As a consequence of Theorem 3.18 of \cite{LXr} we have

\bprop\label{318}
 $$ S(\omega, \omega_1\otimes \omega_2)=
\lim_{p\rightarrow 1} \Tr(\sigma_{C_p})=\Tr(\sigma_C)
$$ where $p\rightarrow 1$ strongly.
\eprop

\section{Estimation of relative entropy}\label{general}
Proposition \ref{318}  suggests that it is useful to study the
following general problem. Let $\H$ be a Hilbert space of countable
dimension, and $P$ be a projection on $\H$. Let $A$ be a positive
bounded operator on $\H$ and $B:= PAP + (1-P)A(1-P).$ It is useful
to note that if $U:= 2P-1$, then $U^2= 1$ and $B= \frac{1}{2}( A+
UAU).$ In particular $B\geq \frac{1}{2}A.$ Let $\tau_A:=PA\ln AP
+(1-P)A\ln A(1-P) - B\ln B.$  Then the problem is to
compute/estimate $\Tr(\tau_A).$

\bprop\label{inc} (1) $$\tau_A = \int_{0}^{\infty} t(P\frac{1}{t+A}P
+ (1-P)\frac{1}{t+A}(1-P)- \frac{1}{t+B}) dt;$$\par (2)
$$\tau_{A+\epsilon} \leq \tau_A, \forall \epsilon >0.$$ \eprop
\proof Ad (1): We note that $||t(P\frac{1}{t+A}P +
(1-P)\frac{1}{t+A}(1-P)- \frac{1}{t+B})||\leq 3,$ and
$$t(P\frac{1}{t+A}P + (1-P)\frac{1}{t+A}(1-P)- \frac{1}{t+B})= P(\frac{B}{t+B}- \frac{A}{t+A})P + (1-P)(\frac{B}{t+B}- \frac{A}{t+A})(1-P),$$ so its  norm is bounded by constant multiplied by $t^2$ when $t$ goes to infinity, hence the improper integral is absolutely convergent in operator norm, and (1) follows by functional calculus.
Ad (2): By (1) we have
\begin{align*}
\tau_{A+\epsilon} & = \int_{0}^{\infty} t(P\frac{1}{t+A+\epsilon}P +
(1-P)\frac{1}{t+A}(1-P)- \frac{1}{t+B+\epsilon}) dt =  \\
& \int_{\epsilon}^{\infty}(t-\epsilon) (P\frac{1}{t+A}P +
(1-P)\frac{1}{t+A}(1-P)- \frac{1}{t+B}) dt
\end{align*}
So
\begin{align*}
& \tau_A - \tau_{A+\epsilon}= \int_{0}^{\epsilon} t(P\frac{1}{t+A}P
+ (1-P)\frac{1}{t+A}(1-P)- \frac{1}{t+B}) dt + \\
& \int_{\epsilon}^{\infty}\epsilon (P\frac{1}{t+A}P +
(1-P)\frac{1}{t+A}(1-P)- \frac{1}{t+B}) dt
\end{align*}
Since $\frac{1}{x}$ is operator convex (cf. \cite{Cal}),
$P\frac{1}{t+A}P + (1-P)\frac{1}{t+A}(1-P)- \frac{1}{t+B}\geq 0$ and
(2) is proved. \qed

As a consequence of (2) of the above Proposition, we have the following improvement of Theorem  3.12 of \cite{LXr}:
\bprop\label{imp}

Let $p$ be  a finite rank projection commuting with $P$, and $A,B$
as above.  Assume that $A-B$ is trace class. Then
$$
\Tr (\tau_A) \geq \Tr \big(\tau_{A_p}) \ .
$$
\eprop \proof When $A\geq \epsilon >0$ the proposition is Theorem
3.12 of \cite{LXr}. Now replace $A$ by $A+\epsilon$ and use (2) of
Proposition \ref{inc}, we have $\Tr(\sigma_{A+\epsilon})\leq
\Tr(\sigma_A).$ On the other hand since $\sigma_{A+\epsilon}$
converges to $\sigma_A$ strongly, it follows that
$$\lim_{\epsilon\rightarrow 0} \Tr(\sigma_{A+\epsilon})\geq
\Tr(\sigma_A)$$ and so we have

$$\lim_{\epsilon\rightarrow 0} \Tr(\sigma_{A+\epsilon})= \Tr(\sigma_A)$$  and the proposition follows by
 Theorem  3.12 of \cite{LXr}.
 \qed

As an application of Proposition \ref{imp},  we specialize $A$ and
$\H$ as follows. We take $\H= L^2 (O, {\Bbb C})$, $O=O_1 \cup O_2
\in \PO.$ $P$ is the projection onto $L^2 (O_1, {\Bbb C})$, and $A$
is given by a kernel $K(x,y)=K(x-y)$ which is singular when $x=y$
but smooth when $x\neq y$.

It is instructive to review how $S(\omega,\omega_1\otimes\omega_2)=
\Tr(\sigma_C)$ is proved. Choose a sequence of finite rank
projections $p_n$ which converges strongly to $1$ and commute with
$P$. Then by the property of relative entropy
$S(\omega,\omega_1\otimes\omega_2)= \lim_n \Tr(\sigma_{C_{p_n}}).$
Since $C_{p_n}\geq 0$ converges to $C$ strongly, we have
$S(\omega,\omega_1\otimes\omega_2)\geq \Tr(\sigma_C).$ The reversed
inequality would follow from Theorem 3.12 of \cite{LXr} if we can
drop the assumption that $A$ is strictly positive. In \cite{LXr} we
use regularized kernel and explicit form of resolvent in the chiral
free fermion case (cf. (1) of Lemma 3.15 in \cite{LXr}) to prove the
reversed inequality. Now with Proposition \ref{imp}  we will always
have $S(\omega,\omega_1\otimes\omega_2)= \Tr(\sigma_C)$ even without
knowing the explicit form of the resolvent of $C$. In particular
this identity is also true for free massive fermions, where the
corresponding covariance operator $C$ is given by formula $187$ in
\cite{Casf}.

To motivate the next result, note that  our goal is to estimate
$A\ln A- B\ln B$ when $A-B$ is trace class. The derivative of  $x\ln
x$ is singular at $x=0$, this explains that when $A,B$ has $0$ in
their spectrum one needs additional conditions. Note that the
derivative of $x^2 \ln x $ is bounded when $x$ is close to zero, and
when $A, B$ are positive we can consider $A\ln \sqrt{A} - B\ln
\sqrt{B}$ with condition that $\sqrt{A} - \sqrt{B}$ being trace
class. It is more convenient in applications to replace last
condition by $|{A} -{B}|^{1/2}$ being trace class, and that is the
condition we impose in Theorem \ref{key}.

\blem\label{ba}
$$
||\frac{1}{t+B} A||\leq \frac{||A||^{1/2}}{t^{1/2}}, \forall t>0
$$
\elem
\proof
For any unit vector $\phi\in \H$ we have
$$
|| \frac{1}{t+B} A \phi||^2 = \langle A \frac{1}{(t+B)^2} A \phi,\phi\rangle$$

Note that $(t+B)^2= t^2 + 2tB +B^2\geq t^2+ t A,$ and so $$ A
\frac{1}{(t+B)^2} A \leq  A \frac{1}{t^2+ tA} A $$ and
$$
\langle A \frac{1}{(t+B)^2} A \phi,\phi\rangle \leq
\frac{1}{t}\langle A \frac{1}{t+A} A \phi,\phi\rangle \leq ||A||
\frac{1}{t}$$ and the Lemma follows. \qed

\bthm\label{key} Suppose that $|A-B|^{\frac{1}{2}}$ is trace class,
then $\tau_A$ is also trace class. \ethm \proof

By (1) of Proposition \ref{inc}, $$\tau_A = \int_{0}^{1}
t(P\frac{1}{t+A}P + (1-P)\frac{1}{t+A}(1-P)- \frac{1}{t+B}) dt +
\int_{1}^{\infty} t(P\frac{1}{t+A}P + (1-P)\frac{1}{t+A}(1-P)-
\frac{1}{t+B}) dt $$ Note that $$\int_{1}^{\infty} t(P\frac{1}{t+A}P
+ (1-P)\frac{1}{t+A}(1-P)- \frac{1}{t+B} dt = -B\ln (B +1) + PA\ln
(A +1)P + (1-P) A\ln (A +1) (1-P) $$  By Lemma 3.11 of \cite{LXr}
$\ln (A+1)- \ln (B+1)$ is trace class, it follows that
$$A\ln (A+1) - B\ln (B+1) = A (\ln (A+1) -\ln (B+1)) + (A-B) \ln (B+1) $$ is trace class, and so is $-B\ln (B +1) + PA\ln (A +1)P + (1-P) A\ln (A +1) (1-P)$. Hence it is sufficient to show that
$$\int_{0}^{1} t(P\frac{1}{t+A}P + (1-P)\frac{1}{t+A}(1-P)- \frac{1}{t+B}) dt$$ is trace class. Let $0< \epsilon<1$ be a small number, and
denote by
$$D_\epsilon: = \int_{\epsilon}^{1} t(P\frac{1}{t+A}P + (1-P)\frac{1}{t+A}(1-P)- \frac{1}{t+B})dt$$ We note that $D_\epsilon\geq 0$ is an increasing sequence of
positive trace class operators which converge in norm to
$$\int_{0}^{1} t(P\frac{1}{t+A}P + (1-P)\frac{1}{t+A}(1-P)-
\frac{1}{t+B}) dt,$$ it is sufficient to show that $\Tr (D_\epsilon)
$ is bounded by a constant independent of $\epsilon$.

By assumption $B-A$ is trace class, we can find an ONB of $\psi_i$ of $\H$ which are the eigenvectors of $B-A$  with eigenvalues $\lambda_i$.
We have
\begin{align*}
\Tr (D_\epsilon) = \int_{\epsilon}^{1} \Tr(t(\frac{1}{t+A}  -
\frac{1}{t+B})) dt & = \int_{\epsilon}^{1}
\Tr(t(\frac{1}{t+B}\frac{1}{t+A}(B-A))) \\
& =\sum_{i} \lambda_i \int_{\epsilon}^{1}\langle \frac{t}{t+A}
\psi_i , \frac{1}{t+B}\psi_i\rangle
\end{align*}

where the interchange of sum and integral in the third equality
follows since the integrand is a continuous function of $t$ in
tracial norm. First assume that $\lambda_i >0 $. Then from
$(t+B)\psi_i = A\psi + (t+\lambda_i)\psi_i$ we have
$$
\frac{1}{t+B} \psi_i = \frac{1}{t+\lambda_i} (\psi_i - \frac{1}{t+B}
A \psi_i)
$$
Hence
$$
\langle \frac{t}{t+A} \psi_i , \frac{1}{t+\lambda_i} (\psi_i -
\frac{1}{t+B} A \psi_i) \rangle \leq \frac{1}{t+\lambda_i}
||\frac{t}{t+A}\psi_i|| (||\frac{1}{t+B} A|| + 1) \leq
\frac{1}{t+\lambda_i} (||A||^{\frac{1}{2}} \frac{1}{t^{\frac{1}{2}}}
+ 1)
$$
where in the last step we have used Lemma \ref{ba}. So
$$
|\int_{\epsilon}^{1}\langle \frac{t}{t+A} \psi_i ,
\frac{1}{t+B}\psi_i\rangle| dt \leq \int_{\epsilon}^{1}
\frac{1}{t+\lambda_i} (||A||^{\frac{1}{2}} \frac{1}{t^{\frac{1}{2}}}
+ 1) dt \leq \pi
||A||^{\frac{1}{2}}\frac{1}{\lambda_i^{\frac{1}{2}}}
+\ln(1+\lambda_i) - \ln \lambda_i
$$
When $\lambda_i <0 $, exchanging the roles of $A,B$ as above we have
$$
\frac{1}{t+A} \psi_i = \frac{1}{t-\lambda_i} (\psi_i - \frac{1}{t+A}
B \psi_i)
$$
We have
\begin{align*}
& \langle \frac{1}{t+A} B \psi_i , \frac{t}{t+B}  \psi_i\rangle
=\langle \frac{1}{t+A} (A+\lambda_i) \psi_i, \frac{t}{t+B}
\psi_i\rangle
\\
& =\langle \frac{1}{t+A} A \psi_i, \frac{t}{t+B}   \psi_i\rangle +
\langle \frac{t}{t+A}  \psi_i, \frac{1}{t+B}  (B-A) \psi_i\rangle
\end{align*}

and it follows that
$$
|\langle \frac{1}{t+A} B \psi_i, \frac{t}{t+B}  \psi_i\rangle| \leq
2 + ||\frac{1}{t+B}  A|| \leq 2+ ||A||^{\frac{1}{2}}
\frac{1}{t^{\frac{1}{2}}}
$$
It follows that
$$
|\int_{\epsilon}^{1}\langle \frac{t}{t+A} \psi_i ,
\frac{1}{t+B}\psi_i\rangle| dt \leq \int_{\epsilon}^{1}
\frac{1}{t-\lambda_i} (3+ ||A||^{\frac{1}{2}}
\frac{1}{t^{\frac{1}{2}}}) dt  \leq \pi
||A||^{\frac{1}{2}}\frac{1}{\lambda_i^{\frac{1}{2}}} +
3(\ln(1-\lambda_i) - \ln (-\lambda_i))
$$

Putting these two cases together we have
$$
\Tr (D_\epsilon) \leq \sum_i |\lambda_i| (  \pi
||A||^{\frac{1}{2}}|\lambda_i|^{\frac{1}{2}} + 3 (\ln
(1+|\lambda_i|)- \ln( |\lambda_i|) )
$$
By assumption $\sum_i |\lambda_i|^{\frac{1}{2}} < \infty$, it follows that $
\Tr (D_\epsilon)$ is bounded by a number independent of $\epsilon$ and the proof is complete.

\qed

To apply Theorem \ref{key} to the computation of relative entropy in
free QFT, we specialize $A$ and $\H$ as follows. We take $\H= L^2
(O, {\Bbb C})$, $O=O_1 \cup O_2 \in \PO.$ $P$ is the projection onto
$L^2 (O_1, {\Bbb C})$, and $A$ is given by a kernel $K(x,y)=K(x-y)$
which is singular when $x=y$ but smooth when $x\neq y$.

\blem\label{fan}
(1) Suppose  $F_1 = PF(1-P) + (1-P) F P$ where $P$ is a projection, and $|F|^{\frac{1}{2}}$ is trace class, then  $|F_1|^{\frac{1}{2}}$ is also trace class, and $|PFP|^{1/2}$ is also trace class.

\elem \proof Let $U=2p-1$. Then $F_1= \frac{1}{2} (F - UFU).$ For a
compact operator $T$, we shall denote by $\mu_n(T)$  its $n$-th
largest singular value among all nonzero eigenvalues. By Fan's
Theorem (cf. \S1 of \cite{Sim}), we have
$$
\mu_{n+m+1}(F-UFU)\leq \mu_{n+1}(F) + \mu_{m+1}(UFU) =\mu_{n+1}(F) + \mu_{m+1}(F), \forall n,m \geq 0$$

Choose $n=m\geq 0$ we have $\mu_{2n+1}(F-UFU)\leq 2\mu_{n+1}(F)$,
and choose $n=m+1$ with $n\geq 1$ we have $$\mu_{2n+1}(F-UFU)\leq
\mu_{n+1}(F) +\mu_n (F).$$ It follows that
$(\mu_{2n+1}(F-UFU))^{1/2}\leq 2^{1/2}\mu_{n+1}(F)^{1/2},
(\mu_{2n}(F-UFU))^{1/2}\leq \mu_{n+1}(F)^{1/2} + \mu_{n}(F)^{1/2}$,
and so $\Tr (|F_1|^{1/2}) = \sum_n \mu_n(F_1)^{1/2} \leq
(\sqrt{2}+1) \Tr (|F|^{1/2})$ . The same argument with $UFU$
replaced by $-UFU$ shows that $|F+UFU|^{1/2}$ is trace class. Note
that $|PFP|\leq \frac{1}{2}|F + UFU|$, the second statement in the
Lemma also follows. \qed

\bprop\label{smooth} Suppose $\H= L^2 (O, {\Bbb C})$, $O=O_1 \cup
O_2 \in \PO, $ $P$ is the projection onto $L^2 (O_1, {\Bbb C})$, and
$A$ is given by a kernel $K(x,y)=K(x-y)$ which is singular when
$x=y$ but smooth when $x\neq y$. Then $|A-B|^{1/2}$ is trace class.
\eprop

\proof

By assumption $A-B$ is given by a kernel $K(x,y)= K(x-y) $ which is
a smooth function for $x\in O_1, y\in O_2$. Choose a large cube $CU$
of length $L$ centered at origin whose interior  contains the
closure of the union of $O_1, O_2, O_1-O_2, O_2-O_1$, and let
$G(x_1,..., x_d)$ be a smooth function on $CU$ such that $G(x-y)=
K(x-y)$ whenever $x\in O_1, y\in O_2$, $\bar G(x)= G(-x)$ and $G$ is
periodic in each of its variables with period $L$. The operator $T$
on $L^2 (CU, {\Bbb C})$ given by kernel $G(x-y)$ can be diagonalized
by Fourier transformation, and its eigenvalues as  functions of
$(n_1,... n_d)$ where $n_i$ are integers go to zero faster than the
inverse of any polynomial in $n_1,... n_d$. It follows that
$|T|^{1/2}$ is trace class. $A-B$ is $P F (1-P) + (1-P) F P$, where
$F$ is the restriction of $T$ to subspace $L^2 (O_1 \cup O_2, {\Bbb
C}).$ By Lemma \ref{fan} the Proposition is proved.

\qed

Since the computation of relative entropy in quantum field theory of
free fermions reduces to finite linear combinations of traces of
$\tau_A$ where $A$ is as in Proposition \ref{smooth}, combined with
Theorem \ref{key} we have proved the following:

\bcor\label{free} The mutual information in quantum field theory of
free fermions  on  Minkowski spacetime of any dimension is finite.
\ecor

\begin{remark}\label{bos} For free boson case there is a formula (cf.
equation $(63)$ of  \cite{Casf}) for mutual information , but the
corresponding operator $C$ there is unbounded and  does not seem to
have a good kernel representations. In the case of chiral massless
free bosons there has been recent computation of mutual information
in the case of two intervals with diagolization of a non-symmetric
operator (cf. \S3 of \cite{Casb}). We note that in the later case
the mutual information is finite since it is embedded into free
fermions. It is an interesting question to see if one can have a
similar treatment of free bosons in general cases as in this
section. \end{remark}

\section{Chiral CFT}\label{cft}
We shall refer the reader to \S2 of \cite{LXr} for the definition of
conformal net and its properties.

A positive lattice $L$ of rank $n$ is the $\Bbb Z$ span of a basis $\alpha_1, ..., \alpha_n$ in a vector space with a positive definite inner product $\langle, \rangle$
 such that $\langle \alpha_i, \alpha_j \rangle \in {\Bbb Z}, \forall 1\leq i,j\leq n.$ $L$ is called even
 if $\langle \alpha_i, \alpha_i \rangle \in 2{\Bbb Z}, \forall 1\leq i\leq n.$ To each even positive lattice $L$ one can associate a rational conformal net $\A_L$ (cf. \cite{DX}).
 The free fermion net $\A_r$ can be considered as conformal net associated with ${\Bbb Z}^r$ with its usual Euclidean inner product.  $\A_r$ is not local, but graded local since ${\Bbb Z}^r$ is not even.

\blem\label{tri} Let $L$ be a  positive lattice  with a basis
$\alpha_1, ..., \alpha_n,$ and for $k$ a positive integer, let $kL$
be the  $\Bbb Z$ span of a basis $k\alpha_1, ..., k\alpha_n.$ Then
$\A_{kL}\subset \A_L$ has index $n^k$. \elem \proof By \cite {DX}
the vacuum representation of $\A_L$ decomposes into finitely many
irreducible representations  of $\A_{kL}$, which are in one to one
correspondence with abelian group of $L/kL,$  and all such
representations have index $1$.  Note  $L/kL$  is isomorphic to
direct product of $n$ copies of ${\Bbb Z}/k{\Bbb Z},$ the Lemma
follows. \qed

\bprop\label{lat} Let $L$ positive lattice $L$ of rank $n$ with a
basis $\alpha_1, ..., \alpha_n.$ Then: \par
 (1)There exists a ${\Bbb Z}$
linear injective map $\phi: L\rightarrow {\Bbb Q}^r$ for some
positive integer $r$ such that $\phi$ preserves inner product ;\par
(2) There exists a positive integer $k$ such that the image of $kL$
under $\phi$ lies in ${\Bbb Z}^r$.

 \eprop \proof Ad (1) It is equivalent  to show that for some positive
integer $r$ there exist vectors $A_i = (A_{1i}, ..., A_{ri}) \in
{\Bbb Q}^r$ such that
$$\sum_{1\leq k\leq r} A_{ki}A_{kj}= \langle \alpha_i, \alpha_j\rangle, \forall 1\leq i,j\leq n.$$ We prove this by induction on $n$.
When $n=1$, one can take $A_1 = (1,...,1)$ with $r= \langle
\alpha_1, \alpha_1\rangle.$ Assume that the Proposition is true for
$n-1$, i.e., for some positive integer $r$ there exist vectors $A_i
= (A_{1i}, ..., A_{ri}) \in {\Bbb Q}^r$ such that
$$\sum_{1\leq k\leq r} A_{ki}A_{kj}= \langle \alpha_i, \alpha_j\rangle, \forall 1\leq i,j\leq (n-1).$$

First we choose a vector $\tilde {A_n}$ in the linear span of $A_1,
..., A_{n-1}$ such that $\langle \tilde {A_n}, A_i\rangle= \langle
\alpha_n, \alpha_i\rangle, \forall 1\leq i\leq n-1.$ Suppose that $
\tilde {A_n} = \sum_{1\leq i\leq n-1} x_i A_i$, then we have a
system of linear equations
$$ \sum_{1\leq j\leq n-1} x_j \langle \alpha_j, \alpha_i\rangle = \langle \alpha_n, \alpha_i\rangle, 1\leq i\leq n-1.$$ Since  $\langle \alpha_j, \alpha_i\rangle$
are integers, it follows that $x_i \in {\Bbb Q}.$ Moreover, we note
that  $ \tilde {\alpha_n} := \sum_{1\leq i\leq n-1} x_i \alpha_i$ is
exactly the projection of $\alpha_n$ onto the linear space spanned
by $\alpha_i, 1\leq i\leq n-1,$  and it follows that
$$ \langle \tilde \alpha_n, \tilde\alpha_n\rangle = \sum_{1\leq k\leq r} \tilde{A}_{kn}  \tilde{A}_{kn}\in  {\Bbb Q}$$
Since $\langle \tilde \alpha_n, \tilde \alpha_n\rangle < \langle
\alpha_n, \alpha_n\rangle,$ we have
$$ \langle  \alpha_n, \alpha_n\rangle- \sum_{1\leq k\leq r} \tilde{A}_{kn} \tilde{A}_{kn} = \frac{p}{q} $$
with both $p,q$ positive integers. Let $s= r+ pq$  and  $A_n$ be a
vector in ${\Bbb Q}^s$ whose first $r$ entries are that of $ \tilde
{A_n},$ and the last $pq$ entries are all $\frac{1}{q}$, and we
embed $A_i$ into ${\Bbb Q}^s$ by simply adding last $pq$ zeros to
the components of $A_i,  1\leq i\leq n-1$, and we have proved the
Proposition for $n$. By induction the proof is complete. \par

Ad (2): The image of each $\phi(\alpha_i)$ has components in ${\Bbb
Q},$ we just have to choose an integer $k$ such that $k$ multiply
each of these components are in ${\Bbb Z}$.

\qed

By Proposition \ref{lat} we immediately have:

\bcor\label{emb} Let $L$ be an even positive lattice and $\A_L$ the
associated conformal net. Then there exist  positive integers $k, r$
such that $\A_{kL}$ is a subnet of free $r$ fermion net $\A_r$.
\ecor

By Corollary \ref{emb} and Lemma \ref{tri} we have proved the
following:

\bcor\label{lat2}  Let $L$ be an even positive lattice and $\A_L$
the associated conformal net. The mutual information for  $\A_L$ is
finite. \ecor

If $G$ is a simply connected simply laced compact Lie group, it is
known (cf. \cite{PS}) that $\A_{G_1}$ is conformal net associated
with a lattice. When $G$ is not simply laced, $G$ is of type $B, C,
F_4, G_2$. Note that $SO(2n+1)\subset SO(2n+2), G_2\subset
F_4\subset E_6$, and $Sp(n)\subset SO(4n)$. So $G_1$ can always be
embedded into $H$ for some simply laced $H.$

Hence the mutual information for $\A_{G_1}$ is finite by Corollary
\ref{lat2}.

Since $\A_{G_k}$ is a subnet of $k$ tensor product of $\A_{G_1}$, it
follows that  the mutual information for $\A_{G_k}$ is finite by
Corollary \ref{lat2}.  It is also clear  that the same is true for
all conformal nets that can be obtained from  cosets, orbifolds,
simple current extensions or combinations of these constructions
starting with $\A_{G_k}$ or $\A_L$, and of course any subnets of
such conformal nets.

\subsection{A relation between relative entropy and
index}\label{rel}

Fix an interval $I_1=(a,b)$ on a circle. Suppose $\rho$ is an
irreducible representation of a conformal net $\A$ localized on
$I_1$ with finite index  given by $\lambda^{-1}$. Let
  $J_{n}\subset I_1', n\geq 1$ be an increasing sequence of
intervals such that
$$
\bigcup_n J_{n} =I_1',\quad \bar J_n\cap \bar I_1= a
$$

Let $E_n$ be the conditional expectation from $\A(I_1) \vee \A(J_n)$
to  $\rho(\A(I_1)) \vee \A(J_n)$. Clearly $E_{n+1}$ is an extension
of $E_n$.

We note that by Pimsner-Popa inequality we have $E_n( a) \geq
\lambda a, \forall $ positive $a\in \A(I_1) \vee \A(J_n).$ Denote by
$\phi_n = \omega E_n.$


\bthm\label{index}
$$
\lim_{n\rightarrow \infty} S(\omega, \phi_n)= \ln (\lambda^{-1})
$$ where $\lambda^{-1}$ is the index of representation $\rho$.
\ethm

\proof

By Section 3.6.2 of \cite{LXr} (We note that there is typo in the
formula in Th. 3.24 of \cite{LXr}, there should be a natural log on
the righthand side) it is sufficient to prove the following as in
Proposition 3.25 of \cite{LXr}: Given any $\epsilon>0$, we need to
find $e\in \A(I_1) \vee \A(J_n)$, such that
$$ |\omega (e)-1|< \epsilon, |\omega (e^*)-1|< \epsilon,\ |\omega
(e^*e)-1|< \epsilon,\ |\phi_n (ee^*)-\lambda|< \epsilon \ .
$$

Let $e_1\in \A(I_1)$ be the Jones projection for
$\rho(\A(I_1))\subset \A(I_1)$, and $v\in \rho(\A(I_1)$ be the
isometry such that $\lambda v^* e_1 v= 1.$ Since $\rho$ is
irreducible, $\rho(\A(I_1))\vee \A(I_1') = \B(\H).$ And since
$\vee_n \A(J_n) =\A(I_1'),$ we can find a sequence of elements
$e_n\in \rho(\A(I_1)) \vee \A(J_n), n\geq 2 $ which converges in
strong star topology to $e_1$.  Now choose $ x_n = \lambda^{-1} v^*
e_1 e_n v.$ Then $x_n \rightarrow 1$ in strong star topology , and
so $\omega (x_n), \omega (x_n^*x_n) $ converges to $1$. On the other
hand by definition
$$
E_n (x_n^* x_n) = v^* e_n^* e_n v
$$ converges to $v^*e_1 v = \lambda^{-1}$ strongly. Hence given  any
$\epsilon>0$, we can choose  $n$ sufficiently large such that if we
set $e=x_n$, then  $e\in \A(I_1) \vee \A(J_n)$, and
$$ |\omega (e)-1|< \epsilon, |\omega (e^*)-1|< \epsilon,\ |\omega
(e^*e)-1|< \epsilon,\ |\phi_n (ee^*)-\lambda|< \epsilon \ .
$$

\qed

One way of interpreting Theorem \ref{index} is the following: Let $
I\subset I_1\cup \bar J_n$£¬ and denote by  $\omega_I, \phi_{n,I}$
the restriction of  $\omega, \phi_n$ to $\A(I)$ respectively. When
$I\subset I_1\cup \bar J_n$ is disjoint from $I_1,$ by definition
$\omega= \phi_n$ and $S(\omega_I, \phi_{n,I}) =0$. Then as $I$
increases,  $S(\omega_I, \phi_{n,I})$ will increase. When $I=I_1\cup
\bar J_n$ increases so that $\vee_n J_n =I_1',$ Theorem \ref{index}
states that the limiting value is $\ln (\lambda^{-1}).$ This picture
has some similarity (but not the same )to the result in \cite{HNT}.

\section{Outlook}\label{out}
There are a number of questions which come naturally from our work.
Does split property imply finiteness of mutual information? Or less
strongly does nuclearity (cf. \cite{BAL}, \cite{Nar}) imply
finiteness of mutual information? Having established finiteness of
mutual information in Corollary \ref{free} the next step will be to
understand the singularity structure when the distance between
regions goes to zero as in \S4.2 of \cite{LXr}, and determine the
regularized entropy. These may be related to the heuristic
computations in \S3 of \cite{Casf}, and may provide rigorous
definition of $C$ function (cf. \cite{Casc}) starting with cut off
independent relative entropy as   in \S4.2 of \cite{LXr} for CFT.
Finally it will also be interesting to determine the asymptotics of
the  mutual information when the distance between regions goes to
infinity and compare with the heuristic computations in \cite{Car}.
We plan to address some of these questions in the future.

\noindent {\bf Acknowledgements}. Part of this work was done when I
participated in Pitp 2018 at Institute for Advanced Study. I  would
like to thank E. Witten for making my visit possible and stimulating
discussions, and R. Longo for helpful comments.

{\footnotesize

\end{document}